\documentclass[prd,showkeys]{revtex4}
\usepackage{caption}
\usepackage{amsmath}
\usepackage{graphicx}
\usepackage{epsfig}
\usepackage{dcolumn}
\usepackage{bm}
\usepackage{slashed}
\usepackage{amsfonts}
\usepackage{amssymb}

\begin{document}

\title{Speed addition and closed time cycle in Lorentz-non-invariant theories%
}
\author{A.E. Shabad}
\email{shabad@lpi.ru}
\affiliation{\textsl{P. N. Lebedev Physics Institute, Leninsky Prospekt 53, Moscow \
117924, Russia} \\
\textsl{Tomsk State University, Lenin Prospekt 36, Tomsk 634050, Russia} }

\begin{abstract}
In theories, whose Lorentz invariance is violated by involvement of an
external \ any-rank tensor, we show that the standard relativistic rule
still holds true for summing the signal speed, understood as the group
velocity of a wave, with the speed of the reference frame. Provided a
superluminal signal is available, this observation enables one to arrange a
closed time cycle and hence causality violation, notwithstanding the Lorentz
noninvariance. Also an optical anisotropy of a moving medium, isotropic at
rest, is revealed.
\end{abstract}

\maketitle

\section{Introduction}

\subsection{About covariant description of Lorentz-non-invariant theories}

In recent years much attention has been paid to the models, wherein the
relativistic invariance does not take place. In such formulation this trend
was initiated in \cite{Kostelecky}. It is aimed at detection of possible
small deviations from the Standard Model, manifesting a violation of the
Special Theory of Relativity. The object of investigation dealt with in
those works is a special four-rank tensor peculiar to the vacuum, and not
formed by the current fields. A special case of a Lorentz-non-invariant
theory beyond the Standard Model is provided by space-time noncommutative
theories (see the review articles \cite{NC, NC2}). On the other hand, well
within the Standard Model, we often consider theories where the invariance
with respect to the Lorentz boosts, as well to the spatial rotations, is
destroyed by the presence of external fixed tensors. These may be due to the
presence of a medium or/and to external fields or/and to nontrivial metrics.
In the first case the role of the external tensor, first-rank in this
instance, is played by the four-velocity vector of the medium. In other
cases these are the tensors of external fields or the ones characterizing
the external metrics, also the noncommutativity tensor, when a
non-commutative theory is under consideration.

In all these cases the theory may be given a Lorentz-covariant form. Namely,
the action is formed as a Lorentz-scalar combination of fields and external
tensors. So are the generating functionals of the Green functions and of the
vertices (see, $e.g$., \cite{Weinberg}). The Green functions and the
one-particle-irreducible vertices -- second- and higher-rank polarization
tensors -- obtained by differentiation of the above functionals over the
vector currents and fields, respectively, are formed as matrices constructed
using the external tensors, apart from particle momenta or coordinates. In
quantum electrodynamics with external electric and magnetic fields such
program was realized in \cite{Batalin} (see also \cite{trudy}). Perhaps, the
first example of covariant representation of the photon polarization tensor
in a (moving) medium making use of its four-velocity may be found in \cite%
{Fradkin}. For the case of a medium in a magnetic field see \cite{Perez},
\cite{trudy}. (A microscopic foundation of such a treatment of the
four-velocity was given in \cite{Cabo} based on the temperature Green
function formalism, see more comments in Subsection \ref{IIC}.) The
covariant treatment of the noncommutativity matrix as a second-rank tensor
was used with the same purpose in \cite{Fresneda}, where also the third-rank
polarization tensor in external electric and magnetic fields responsible for
the nonlinear processes of photon splitting and merging are built in the
same manner. The covariant representation for the third-rank polarization
tensor in a medium was used in \cite{Ferrer}.

The covariant approach is as a matter of fact a realization of the evident
"extended relativity principle" to be read as: An inertial Lorentz reference
frame A\ with an external tensor in it is equivalent to another reference
frame B that moves with respect to A with a constant speed $\mathbf{V}$,
provided that the external tensor in the frame B is the one, which has been
Lorentz-transformed from that in A with the help of the speed $\mathbf{V.}$

The gauge invariant energy-momentum tensor in this approach is not
symmetric, its antisymmetric part being responsible for nonconservation of
the generators of angular momentum and Lorentz boosts \cite{Chavez}.

In the present paper, following this approach we establish the
transformation law of the group velocity of electromagnetic wave packet from
one inertial frame to another to show that it is just the standard rule of
relativistic summation of the two speeds: the group velocity of the packet
and the relative speed of the two inertial reference frames.

We are studying the group velocity, since this quantity is the speed of
propagation of a wave packet \cite{bornwolf}, and it is significant for the
issues of causality, discussed in Section \ref{III}, although some authors
often refer to the "phase" velocity, what is not grounded in this
connection. The only reservation, usually made about the group velocity, is
that it may exceed the speed of light near a resonance, causing the
so-called abnormal dispersion. However, this reservation, too, can be lifted
\cite{2010} by considering the real part of the complex group velocity as a
function of real momentum, and not real group velocity plotted against the
real part of complex momentum. As for the velocity of the wave front,
referred to as the signal speed after Brillouin \cite{bril}, it always
propagates with the speed of light in the vacuum, since it is determined by
the infinite-frequency limit $k_{0}=\infty ,$ whereas in this limit no
effect of Lorentz-violating factors, like an external field or a medium, can
survive. Consequently, this speed is not important in discussing the
causality issues.

The main derivation is given in Subsection \ref{IIB}, preceded by the
description of the covariant approach in Subsection \ref{IIA} and followed
by consideration of the interesting special case of light propagation in a
moving medium, discussed as an illustration in Subsection \ref{IIC}. It is
shown that the dielectric tensor of an isotropic medium is anisotropic in a
moving frame, the principle axes being fixed by the directions of the
wave-vector and the velocity $\mathbf{V.}$ In the concluding Section \ref{III}%
, among other issues of causality, we are discussing the arrangement
of a closed time cycle in a Lorentz-non-invariant theory.

\section{Lorentz transformation of the group velocity}

\subsection{Polarization operator and dispersion equations\label{IIA}}

The (second pair of) Maxwell equations linearized above a homogeneous
background for a free electromagnetic wave with the 4-vector potential $%
A^{\rho }(k)$ can be written in the momentum $k_{\mu }$ representation in
the form
\begin{equation}
(k^{2}g_{\mu \rho }-k_{\mu }k_{\rho })A^{\rho }(k)-\Pi _{\mu \rho
}(k)A^{\rho }(k)=0,  \label{ME}
\end{equation}%
where $\Pi _{\mu \rho }(k)$ is the polarization operator defined in the
configuration space as
\begin{equation}
\Pi _{\mu \tau }(x,y)=\left. \frac{\delta ^{2}\Gamma }{\delta A_{\mu
}(x)\delta A_{\tau }(y)}\right\vert _{F=\overline{F}=\mathrm{const}}.
\label{PO}
\end{equation}%
via the effective action $\Gamma $, the generating functional of
one-particle-irreducible vertices, the polarization operator being one of
them \cite{Weinberg}. The effective action $\Gamma $ here is admitted to
depend, apart from external tensors, on any order space-time derivatives of
the electromagnetic field tensor $F_{\mu \nu }$, but the electromagnetic
fields are set constant $\overline{F}_{\mu \nu }$(or zero, in the no
external field case) after the differentiations. The Greek indices above
span the Minkowski space.

Let us present the polarization operator in a diagonal form
\begin{equation}
\Pi _{\mu \tau }(k,p)=\delta (k-p)\Pi _{\mu \tau }(k),\quad\Pi _{\mu \tau
}(k)=\sum_{a=1}^{3}\varkappa _{a}~\frac{\flat _{\mu }^{(a)}~\flat _{\tau
}^{(a)}}{(\flat ^{(a)})^{2}},  \label{diag}
\end{equation}%
where $\flat _{\tau }^{(a)}$ are its eigenvectors
\begin{equation*}
\Pi _{\mu \tau }~\flat _{\tau }^{(a)}=\varkappa _{a}~\flat _{\mu
}^{(a)},\quad a=1,2,3,4.
\end{equation*}%
The effective action is a Lorentz scalar formed using all the external
tensors and fields, and it does not depend on coordinates explicitly. Hence,
the polarization operator (\ref{PO}) is a tensor in Minkowski space, while
its eigenvalues $\varkappa _{a}$ are scalars. The appearance of the
energy-momentum conserving delta-function in (\ref{diag}) is owing to the
assumption that only space- and time-independent gauge-invariant
Lorentz-violating agents are considered that do not violate the translation
invariance. The fourth eigenvector is trivial, $\flat _{\mu }^{(4)}=k_{\mu }$%
, so the fourth eigenvalue vanishes, $\varkappa _{4}=0,$ as a consequence of
the 4-transverseness of the polarization operator, $\Pi _{\mu \tau }k_{\tau
}=0$. All eigenvectors are mutually orthogonal, $\flat _{\mu }^{(a)}\flat
_{\mu }^{(b)}\sim \delta _{ab}$, this means that the first three ones are
4-transversal, $\flat _{\mu }^{(a)}k_{\mu }=0$.

The connection between $\Pi _{\mu \tau }$ and the four-rank tensor studied
and attempted to be measured in the approach of Refs.\cite{Kostelecky} was
discussed in \cite{Chavez}.

The scalar eigenvalues $\varkappa _{a}$ may depend on all characteristic
scalars in the theory, including $k^{2}$ and those which are formed by
tensors of any rank peculiar to the medium or to the vacuum, when contracted
with the photon 4-momentum $k_{\mu }$, for instance $k\overline{F}^{2}k$ or $%
k\theta \overline{F}k$ \textit{etc., }where\textit{\ }$\overline{F}_{\mu \nu
}$ stands for external field strength tensor, and $\theta _{\mu \nu }$ for
the non-commutativity tensor.\ Among these external tensors there may be the
4-velocity vector of a uniform medium, if the latter is present. There are
also momentum-independent scalars among arguments of $\varkappa $, like the
external field invariants $\mathfrak{F=-}\frac{1}{4}\overline{F}^{2}$ \ and $%
\mathcal{G=}\mathfrak{-}\frac{1}{4}\overline{F}\,\widetilde{\overline{F}}$,
where $\widetilde{F}$ is the dual electromagnetic tensor, but these are
inessential for the present derivation. The photon dispersion laws for each
of three polarization modes are to be found from the equations
\begin{equation}
k^{2}=\varkappa _{a}(k),\text{ \ \ \ }a=1,2,3,  \label{disp}
\end{equation}%
which are the solvability conditions for Eq.(\ref{ME}). Equations (\ref{disp}%
) determine the frequency $k_{0}$ as a function of the wave-vector
components $k_{i}.$ The eigenvectors $\flat _{\mu }^{(a)}$ serve as 4-vector
potentials for free eigenwaves. There are three eigen-modes, and not two,
because possible massive vector particles are also included into the
propagation equation (\ref{ME}). These may, for instance, be the
electron-positron states (mutually free, or bound into the positronium
atom), with which the photon unites into a mixed polariton state \cite%
{positronium}. Another example of the third polarization degree of freedom,
characteristic of massive vector particles, is supplied by the known
longitudinal modes in a medium, see Subsection IIC for further comments.

The dielectric $\varepsilon _{nj}$ tensor of the anisotropic "medium," to
which the vacuum with the broken Lorentz invariance is equivalent
irrespective of whether the real medium is present or not, is connected with
the polarization tensor components as%
\begin{equation}
\varepsilon _{nj}=\delta _{nj}+\frac{\Pi _{nj}}{k_{0}^{2}},\text{ \ \ }n,%
\text{ }j=1,2,3.  \label{epsilon}
\end{equation}%
The Lorentz-noninvariant vacuum manifests itself as an equivalent
anisotropic "medium", since this tensor (\ref{epsilon}) is different from $%
\delta _{nj}.$ If a homogeneous medium , isotropic in its rest frame, is
present alone in the Lorentz-invariant vacuum (i.e. where no external
tensors besides the 4-velocity of the medium are included), it becomes
anisotropic in a moving frame. (We shall elaborate this point in Subsection
IIC below).

The refraction indices (in every eigenmode) are defined on solutions of
dispersion equations (\ref{disp}) as

\begin{equation}
n_{a}(k_{0},\mathbf{k})=\frac{|\mathbf{k}|}{k_{0}}=\left( 1+\frac{\varkappa
_{a}(k_{0},\mathbf{k})}{k_{0}^{2}}\right) ^{\frac{1}{2}}.
\label{refraction index}
\end{equation}%
These are not Lorentz scalars, contrary to the eigenvalues $\varkappa _{a}.$

\subsection{Derivation of the addition rule for speeds\label{IIB}}

Denote the momentum-dependent Lorentz invariants, $k^{2}=$ $%
-k_{0}^{2}+k_{i}^{2}$ included, as $I_{s}$, $s=1,2,..$, their number
depending on the problem. The derivative $\partial I_{s}/\partial k_{\mu
}=P^{(s)\mu }$ is a Lorentz-vector. The group velocity of the photon is the
3-vector defined as the frequency differentiated over the wave vector
\begin{equation}
v_{i}^{\mathrm{gr}}=\frac{\partial k_{0}}{\partial k_{i}},\qquad i=1,2,3
\label{group-1}
\end{equation}%
and calculated on a solution of the dispersion equation (\ref{disp}). We
shall henceforward omit the indexing of the photon modes $a$ in
understanding that our derivation relates to any of the three. We find that
(summation over all invariants $I_{s},$ whatever their number may be is
meant)
\begin{equation}
v_{i}^{\mathrm{gr}}=-\left( \frac{\partial (k^{2}-\varkappa )}{\partial k_{i}%
}\right) \left( \frac{\partial (k^{2}-\varkappa )}{\partial k_{0}}\right)
^{-1}=\frac{2k_{i}-X_{s}P_{i}^{(s)}}{2k_{0}-X_{s}P_{0}^{(s)}},
\label{group2}
\end{equation}%
where the quantities $X_{s}=\frac{\partial \varkappa }{\partial I_{s}}$ are
Lorentz-invariant.

Let us now imagine that an inertial Lorentz frame $W$, to which the previous
equations relate, moves with respect to the initial frame $W^{\prime }$ --
to be marked with prime -- with speed $\mathbf{V}$. We are going to
demonstrate that the group velocity $\mathbf{v^{\prime \mathrm{gr}}}$ in the
frame $W^{\prime }$ is connected with that in the frame $W$ by the standard
relativistic rule $\mathbf{v^{\prime }{}=v\oplus V}$ of summing the speed of
a signal $\mathbf{v}$ with the speed of the reference frame, with the group
velocity taken for $\mathbf{v.}$ Namely, we shall demonstrate that
\begin{equation}
v_{\parallel }^{\prime }{}^{\mathrm{gr}}=v_{\Vert }^{\mathrm{gr}}\oplus
\mathbf{V}\equiv \frac{V+v_{\Vert }^{\mathrm{gr}}}{1+Vv_{\Vert }^{\mathrm{gr}%
}},  \label{rulepar}
\end{equation}%
\begin{equation}
\mathbf{v}_{\perp }^{\prime }{}^{\mathrm{gr}}=\mathbf{v}_{\perp }^{\mathrm{gr%
}}\oplus \mathbf{V}\equiv \frac{\mathbf{v}_{\perp }^{\mathrm{gr}}\sqrt{%
1-V^{2}}}{1+Vv_{\Vert }^{\mathrm{gr}}},  \label{ruleperp}
\end{equation}%
where the subscripts $\parallel $ and $\perp $ mark projections to the
directions, parallel and orthogonal to $\mathbf{V.}$\

We accept the view that the same physical process that underlies the signal
propagating with the speed $\mathbf{v}$ in the rest frame $W$ should be
responsible for its propagation with the speed $\mathbf{v}\oplus \mathbf{V}$
in the moving frame $W^{\prime }$ following equations of motion for the
signal propagation covariantly transformed from $W$ \ to $W^{\prime }$.
Namely, we adopt that the physical carrier of a signal is an electromagnetic
wave process governed by equations (\ref{ME}), and that the role of the
signal speed will be played by the group velocity of a propagating packet.
We are essentially basing on the fact established in the previous
Subsection, that \ the right-hand side $\varkappa $ of the dispersion
equation (\ref{disp}) is a Lorentz-invariant, to efficiently account for the
mentioned Lorentz transformation of the electromagnetic signal speed.

Using eq.(\ref{group2}) in eq.(\ref{rulepar}) we get
\begin{eqnarray}
v_{\Vert }^{\mathrm{gr}}\oplus v &=&\left( V+\frac{2k_{\Vert }-X_{s}P_{\Vert
}^{(s)}}{2k_{0}-X_{s}P_{0}^{(s)}}\right) \left( 1+V\frac{2k_{\Vert
}-X_{s}P_{\Vert }^{(s)}}{2k_{0}-X_{s}P_{0}^{(s)}}\right) ^{-1}=  \notag \\
&=&\frac{V(2k_{0}-X_{s}P_{0}^{(s)})+2k_{\Vert }-X_{s}P_{\Vert }^{(s)}}{%
2k_{0}-X_{s}P_{0}^{(s)}+V(2k_{\Vert }-X_{s}P_{\Vert }^{(s)})}.
\label{rulepar1}
\end{eqnarray}%
Bearing in mind that the Lorentz transformations for the vectors $\partial
I_{s}/\partial k_{\mu }=P^{(s)\mu }$ (and analogous transformations for the
momenta $k^{\mu }$) between the two frames are
\begin{equation}
P{}_{\Vert }^{\prime }=\frac{P_{\Vert }+VP_{0}}{\sqrt{1-V^{2}}},\qquad
P_{0}^{\prime }=\frac{P_{0}+VP_{\Vert }}{\sqrt{1-V^{2}}},  \label{lorentz}
\end{equation}%
we find from (\ref{rulepar1})
\begin{equation}
v_{\Vert }^{\mathrm{gr}}\oplus \mathbf{V}=\frac{2k_{\Vert }^{\prime
}-X_{s}P_{\parallel }^{\prime }{}^{(s)}}{2k_{0}^{\prime }-X_{s}P^{\prime
}{}_{0}^{(s)}},  \label{rulepar3}
\end{equation}%
which is just $v^{\prime }{}_{\Vert }^{\mathrm{gr}}$, the (parallel
projection of) the group velocity (\ref{group2}) calculated in the frame $%
W^{\prime }$, taking into account that $X_{s}$ are Lorentz-invariant. Thus,
eq.(\ref{rulepar}) is proved.

Analogously, substituting eq.(\ref{group2}) into eq.(\ref{ruleperp}) we
obtain
\begin{eqnarray}
\mathbf{v}_{\perp }^{\mathrm{gr}}\oplus \mathbf{V} &=&\left( \frac{2\mathbf{k%
}_{\perp }-X_{s}\mathbf{P}_{\perp }^{(s)}}{2k_{0}-X_{s}P_{0}^{(s)}}\sqrt{%
1-V^{2}}\right) \left( 1+V\frac{2k_{\Vert }-X_{s}P_{\Vert }^{(s)}}{%
2k_{0}-X_{s}P_{0}^{(s)}}\right) ^{-1}=  \notag \\
&=&\frac{(2\mathbf{k}_{\perp }-X_{s}\mathbf{P}_{\perp }^{(s)})\sqrt{1-V^{2}}%
}{2k_{0}-X_{s}P_{0}^{(s)}+V(2k_{\Vert }-X_{s}P_{\Vert }^{(s)})}=\frac{2%
\mathbf{k}^{\prime }{}_{\perp }-X_{s}\mathbf{P}^{\prime }{}_{\perp }^{(s)}}{%
2k^{\prime }{}_{0}-X_{s}P^{\prime }{}_{0}^{(s)}}.  \label{ruleperp2}
\end{eqnarray}%
We have used again the second equation in (\ref{lorentz}) for the
transformation of $P_{0}^{(s)}$ and analogous transformation for $k_{0},$ as
well as the fact that vector components perpendicular to the relative speed $%
\mathbf{V}$ of the two reference frames do not change under the Lorentz
boost along $\mathbf{V}$. By comparing (\ref{ruleperp2}) with (\ref{group2})
we see that the former is just the perpendicular group velocity $v^{\prime
}{}_{\perp }^{\mathrm{gr}}$ as calculated by the observer in the reference
frame $W^{\prime }$. Thus, the relativistic rule of summing speeds for the
perpendicular component (\ref{ruleperp}) has been proven, too.

The derivation of the relativistic law for speed summation above can by no
means be repeated as applied to the phase velocity $v_{\text{ph}}=\frac{k_{0}%
}{|\mathbf{k}|},$ in correspondence with the well-known fact \cite{landlif}
that \ that velocity cannot serve as a signal speed and is not a 3-vector at
all.

\subsection{Moving isotropic medium\label{IIC}}

An important example of the above consideration is provided by an isotropic
medium, whose presence violates the Lorentz invariance. The equations of
electromagnetic field may be given a Lorentz-covariant form by introducing
the vector of four-velocity $u_{\mu },$ $u^{2}=-1$ of the medium, $%
u_{0}=(1-V^{2})^{-^{\frac{1}{2}}},$ $\mathbf{u}=\mathbf{V}(1-V^{2})^{-^{%
\frac{1}{2}}},$ where $\mathbf{V}$ is the velocity 3-vector of the moving
medium. Then the scalars referred to in the previous Subsection are $%
I_{1}=k^{2}$ and $I_{2}=\left( uk\right) ^{2}.$

The introduction of this vector gives the possibility to covariantly treat
linear \cite{Fradkin} and nonlinear \cite{Ferrer} Maxwell equations of an
initially isotropic medium also after it is set at motion with a constant
speed as a whole and, moreover, placed in external field \cite{Perez}. A
microscopic theory justifying these receipts within the temperature Green
function method in relativistic statistics due to \cite{Fradkin}\ may be
found in \cite{Cabo}. This theory is based on writing the density matrix in
the form $\rho =\exp \left[ -\beta \left( uP\right) \right] ,$ where $\beta $
is the (Lorentz-scalar) inverse temperature, and $P$ is the 4-momentum, in
place of the standard expression $\rho =\exp \left[ -\beta H\right] ,$ to
which it reduces in the rest frame of the medium, where $u_{\mu }=(1,0,0,0),$
$\ H=P^{0}$ is the Hamiltonian. Then the thermodynamical potential, playing
the role of the Lagrangian density in plasma, is written as a Lorentz scalar
with the use of the 4-velocity vector.

The extended relativity principle mentioned in Introduction now reads that
any two inertial frames in a medium are equivalent after the speed of the
medium is Lorentz-transformed from one frame to the other. In more sensual
terms this means that an observer in a frame in motion relative to a medium,
say, air, certainly realizes that he/she is moving, already because he/she
may sense the wind. In this respect the usual relativity principle is
violated together with the Lorenz-invariance. But after the motion of the
medium with respect to the observer is excluded by its Lorentz
transformation the situation for him/her returns to be equivalent to that
viewed upon by an observer at rest with respect to the medium.

The most general covariant tensor representation \cite{Fradkin} for the
polarization operator of an isotropic homogeneous medium may be rewritten in
the diagonal form \cite{Ferrer}, \cite{shabad2009} as
\begin{equation}
\Pi _{\mu \nu }(k)=\kappa \sum_{b=1,2}\frac{c_{\mu }^{(b)}c_{\nu }^{(b)}}{%
(c^{(b)})^{2}}+\kappa _{3}\frac{a_{\mu }a_{\nu }}{a^{2}},  \label{Pidiag}
\end{equation}%
where
\begin{equation*}
a_{\mu }=u_{\mu }k^{2}-k_{\mu }(uk),\qquad
a^{2}=k^{2}(k^{2}-(uk)^{2}),\qquad (au)=0.
\end{equation*}%
Vector $c_{\mu }^{(1)}$ is defined as an arbitrary 4-vector normal to the
hyperplane, spanned by vectors $k_{\mu }$ and $a_{\mu }$. Vector $c_{\mu
}^{(2)}\equiv \varepsilon _{\mu \nu \rho \lambda }c_{\nu }^{(1)}a_{\rho
}k_{\lambda }$ is also normal to that hyperplane and, also to $c_{\mu
}^{(1)} $. The four vectors $k_{\mu },\,a_{\mu }$, and $c_{\mu }^{(1,2)}$
are eigenvectors of the polarization operator%
\begin{eqnarray}
\Pi _{\mu }^{~\nu }c_{\nu }^{(1,2)} &=&\kappa _{1,2}c_{\mu }^{(1,2)},\text{
\ }\kappa _{1}=\kappa _{2}=\kappa  \notag  \label{eigena} \\
\Pi _{\mu }^{~\nu }a_{\nu } &=&\kappa _{3}a_{\mu },\qquad \Pi _{\mu }^{~\nu
}k_{\nu }=0.
\end{eqnarray}%
Only three of them are involved in the decomposition (\ref{Pidiag}), since
one eigenvalue is zero, in accord with the transversality $\Pi _{\mu }^{~\nu
}k_{\nu }=0$ .

\ The three basis vectors $\,a_{\mu }$, and $c_{\mu }^{(1,2)}$ are 4-vector
potentials of the electromagnetic eigen-waves. The orientations of the
corresponding electric, $e_{i}\mathbf{\sim }$ $k_{0}a_{i}\mathbf{-}$ $%
k_{i}a_{0}$ (or $e_{i}\mathbf{\sim }$ $k_{0}c_{i}^{(1,2)}\mathbf{-}$ $k_{i}$
$c_{0}^{(1,2}$) and magnetic ($h_{i}=\epsilon _{ijn}k_{j}a_{n}$ or $%
h_{i}=\epsilon _{ijn}k_{j}c_{n}^{(1,2)})$ \ fields, calculated basing on
these vector-potentials, are described in detail in \cite{Ferrer}. In the
Lorentz frame, where the medium is at rest, modes 1 and 2 are
electromagnetic waves transversely-polarized in the plane orthogonal to the
wave vector $\mathbf{k}$, while mode 3 is purely electric wave, its magnetic
field being equal to zero and electric field being longitudinally polarized
along $\mathbf{k}$, $\mathbf{e\sim }$ $\mathbf{k}(k_{0}^{2}-\mathbf{k}^{2})$%
. This wave may be realized provided the dispersion equation has a massive
solution with $(k_{0}^{2}-\mathbf{k}^{2})\neq 0.$ We are saying this with
the only purpose to illustrate, how these two well-known facts appear in our
formalism.

The degeneracy $\kappa _{1}=\kappa _{2}$ in (\ref{eigena}) reflects the
symmetry of the problem under rotations around the direction of the speed $%
\mathbf{V.}$

The dispersion equations (\ref{disp}) take the form%
\begin{equation*}
k^{2}=\kappa _{1,2}(k^{2},(uk)^{2}),\text{ \ \ }k^{2}=\kappa
_{3}(k^{2},(uk)^{2}),
\end{equation*}%
wherein we explicitly indicated the dependence of the eigenvalues on two
Lorentz-scalars $k^{2}$ and $(uk)^{2}$.

In accord with the representation (\ref{Pidiag}), the space-space part of
the polarization tensor, $\Pi _{nj}$, apart from the unit tensor $\delta
_{nj},$ is formed by the wave vector $\mathbf{k}$ and the velocity of the
moving medium $\mathbf{V:}$
\begin{equation}
\Pi _{nj}=A\delta _{nj}+Bk_{n}k_{j}+C\left( k_{n}V_{j}+V_{n}k_{j}\right)
+DV_{n}V_{j},
\end{equation}%
where the coefficients $A$, $B$, $C$ and $D$ are certain rotational scalars,
functions of\textbf{\ }$\mathbf{k}^{2},$ $V,$ $\left( \mathbf{V\cdot k}%
\right) $ and $k_{0}$. Therefore, the vector $\mathbf{d}$\ orthogonal to the
plane spanned by these two vectors, $\left( \mathbf{d\cdot k}\right) =\left(
\mathbf{d\cdot V}\right) =0,$ is a universal eigenvector of the dielectric
tensor (\ref{epsilon})
\begin{equation*}
\varepsilon _{nj}d_{j}=\frac{A}{k_{0}^{2}}d_{i}.
\end{equation*}%
The orientations of the other two eigenvalues in the above plane depend on
individual orientations of the vectors $\mathbf{k}$ \ and $\mathbf{V.}$ This
situation is typical for crystals of monoclinic system \cite{landlif}.

We conclude this Subsection with a statement that an isotropic medium
behaves itself as anisotropic, when it moves. Its dielectric tensor has
three different eigenvalues, all depending on the speed of the medium, and
three optical axes determined by direction of that speed. The refraction
indices depend on the reference frame, too. There are three eigenwaves with
different dispersion laws. Transparent isotropic bodies moving with
relativistic speed are birefringent.

Eqs.(\ref{rulepar}), (\ref{ruleperp}) read that, in a moving frame, the
speed of every eigenwave, considered as its group velocity, is obtained as a
relativistic sum of its group velocity in the rest frame with the speed of
the medium. The same statement concerning a medium, anisotropic already in
the reference frame, where it is at rest, cannot be derived from equations
of the previous Subsection, since it is unclear if 3-dimensional tensors,
responsible for such anisotropy, can be given a relativistic-covariant
extension with the help of the vector $u_{\mu }$.

\section{Concluding remarks\label{III}}

We have demonstrated that when an electromagnetic wave packet is the
physical carrier of information, its group velocity \textbf{\ }obeys the
standard relativistic law (\ref{rulepar}), (\ref{ruleperp}) for speed
summation irrespective of whether the Lorentz invariance of the vacuum holds
true or is violated by the presence of any external vector(s) or any-rank
tensor(s). Thereby, we have shown that the light propagation process that is
governed by electromagnetic field equations depending on that (those)
tensor(s), guarantees that the group velocity in the rest frame and in the
moving frame are related according to the relativistic law of signal's speed
transformation. This indicates that, in the Lorentz-violated vacuum, the
same as in the Lorentz-symmetrical case \cite{bornwolf}, the group velocity
possesses the important property of a signal speed. In particular, when the
group velocity does not exceed $c=1$ (the speed of light in the
Lorentz-invariant vacuum), and the speed of the reference frame does not
exceed unity either, the resulting group velocity in the moving frame
remains lesser than unity. So, normally, the causality survives the
relativistic invariance violation.

However, it sometimes happens that a superluminal signal appears resulting
from dynamical calculations in theories with violated Lorentz-invariance.
The examples are given by light propagation in external metrics \cite%
{Drummond} and noncommutative electrodynamics \cite{Guralnik2001}, \cite%
{Fresneda}, also in QED with an extraordinary strong external field \cite%
{ShabUs2011} and in other systems. (Although in \cite{Guralnik2001} the
phase velocity was used as a criterion for superluminocity, the conclusion
of the authors turns out to be correct, since it relates, as matter of fact,
to the group velocity, as well). There is a temptation to dismiss the grave
character of such results referring to the idea that as long as the
relativistic invariance has been abandoned there in no reason to concern
about causality any longer following the principle expressed in the Russian
proverb: "Sniavshi golovu, po volosam ne plachut!" or in its German
rationalized equivalent: "Ist der Kopf abgeschlagen, wird niemand nach dem
Hute fragen?" This means approximately that the one who is beheaded should
not mourn over his hair or hat. This stand is grounded by the statement \cite%
{Shore} that thanks to the "extended relativity principle" as we were
describing it in Subsection \ref{IIC} one  cannot arrange the closed time
cycle in a Lorentz-noninvariant theory even when a superluminal signal is at
one's disposal. Intuitively, this statement seems paradoxical, because it
admits that, for instance, faster-than-light propagation of acoustic signal
in an infinitely extended perfectly rigid body, whose presence certainly
singles out its reference frame among other frames and hence violates the
relativistic invariance, may be compatible with causality.

We keep, however, to the opposite point of view expressed in \cite%
{dolgovnovikov}; we stress that the causality issues remain
meaningful beyond the Lorenz-invariance, too. Our results allow to
state that when the superluminal signal is carried by the
electromagnetic wave packet in a Lorenz-noninvariant theory, the
causality principle is destroyed in the same way as in a
Lorenz-invariant theory. The point is that the "time machine", i.e.
the closed time cycle, can be constructed in a gedanken experiment
notwithstanding the lack of relativistic invariance. To be more
specific, let a signal be emitted by an  emitter at rest in the
reference frame, designed
as a primed frame in Subsection \ref{IIB}. Let it be emitted in the origin $x'=0$
at the time moment $t/=0$ and then propagate with the superluminal speed $%
v^{\prime }>1$ along the axis 1 to come to a certain point ( $x_{1}^{\prime
},t^{\prime })$ with $x_{1}^{\prime }=v^{\prime }t^{\prime }>t^{\prime }.$
Since this point is space-like, the Lorentz boost exists $x_{1}=$ ($%
x_{1}^{\prime }+Vt^{\prime })/(1-V^{2})^{1/2},$ $t=$($t^{\prime
}+Vx_{1}^{\prime })/(1-V^{2})^{1/2}$ to an (unprimed) frame moving with the
negative underluminal speed $1>$ $|V|>1/v^{\prime }$ that reverses the sign
of the time $t<0.$ To let the signal reflected by a detector at rest in the
unprimed frame come back to the origin and close the time cycle thereby, the
reflected signal should have the speed ${\vert}v|=x/t=$($%
x_{1}^{\prime }+Vt^{\prime })/$($t^{\prime }+Vx_{1}^{\prime }),$
which is just our equation (\ref{rulepar}) for the group velocity in
the moving frame. Now we may perform the inverse boost to come back
to the frame equivalent to the initial primed frame. The origin is
not transformed by this boost.  So, once the signal transmission is
realized via a wave process governed by the Lorenz-invariant (with
external tensors included) dispersion law (\ref{disp}), the
paradoxical influence of a consequence on its cause is achieved
despite the lack of Lorentz invariance. Consequently, situations
where the group velocity turns out to be greater than unity should
be recognized as contradicting the causality principle in a theory
with the Lorentz symmetry violated by an external tensor (as well as
in a Lorentz-invariant theory, of course).

However, it was pointed in Ref. \cite{Fresneda}, where the excess of
the group velocity of an electromagnetic wave in noncommutative
electrodynamics with external constant fields over unity was
analyzed, that its extreme smallness, $v^{\text{gr}}-1\ll 1,$
results in the fact that the speed of the emitter and detector
necessary to realize the paradox of the \textit{influence on the
past }becomes too large, beyond any existing human experience. In
other words, for realizing the closed circle it is not enough to
have a superluminal signal\ at one's disposal, but also some
macroscopical bodies must move with the speed, smaller than, but
very close to that of light, $1>$ $|V|$\textsf{\ }$>1/v.$ The
present-day knowledge does not cover the range of such speeds of
macroscopic instruments. Therefore, deviation from the causality
principle, logically absurd as it is, should not be completely ruled
out on the basis of existing experience, provided there is but a
tiny excess of the signal speed over that of light.

Now we extend that analysis to the superluminal electromagnetic wave
solutions in black hole metrics, discovered within the low-momentum
one-electron-positron-loop approximation in \cite{Drummond}. For the
Schwarzschild metrics the speed excess makes $\Delta v\sim \frac{\alpha }{%
30\pi }\frac{R_{\text{gr}}}{r}\left( \frac{\lambda _{\text{C}}}{r}\right)
^{2},$ where $\alpha =\frac{1}{137}$ is the fine structure constant, $R_{%
\text{gr}}$ is the gravitational radius of the black hole, $\lambda _{\text{C%
}}\sim $ $2.4\times 10^{-10}$cm is the electron Compton wave length, and $r$
is the distance from the center of the star. When the latter is equal to its
gravitational radius, $r=R_{\text{gr}},$ one gets $\Delta v\sim \frac{\alpha
}{30\pi }\left( \frac{\lambda _{\text{C}}}{R_{\text{gr}}}\right) ^{2}.$ Even
for the smallest black holes with their masses of the order of a few solar
masses, where $R_{\text{gr}}\sim 5\times 10^{5}$cm, this excess makes $%
\Delta v\sim 2\times 10^{-35},$ which is much smaller than the
admissible estimates resulting from the noncommutative theory in
\cite{Fresneda} would admit. This means that a tachyon, if emitted
from inside of a black hole, can overcome its horizon with a very
small excess of its speed over the unity to be reduced to nothing as
it goes away from it. On the contrary, deep inside the black hole,
where $r$ is small, the speed excess may be large, only limited by
what the approximations in \cite{Drummond} allow, i.e. $\Delta v\ll
1.$ One can imagine that a tachyon emitted from the depth of the
black hole meets, as it moves toward the horizon, the matter falling
inside with the speed approaching that of light. The latter may
reflect it back towards its source to form the necessary "detail" of
the time machine depicted above. Thus a prospect of developing an
acausal electrodynamics of black holes arises based not only on
retarded, but also on advanced interaction.

\section*{Acknowledgements}

Supported by FAPESP under grant 2014/08970-1, by RFBR under Project
15-02-00293a, and by the TSU Competitiveness Improvement Program, by a grant
from \textquotedblleft The Tomsk State University D.I. Mendeleev Foundation
Program\textquotedblright .

\end{document}